\documentclass[aps,prl,twocolumn,floatfix, showpacs]{revtex4}

\usepackage{graphicx}
\usepackage{epsfig}
\graphicspath{{Figures/}}

\def\gtwid{\mathrel{\raise.3ex\hbox{$>$\kern-.75em\lower1ex\hbox{$\sim$}}}}
\def\alt{\mathrel{\raise.3ex\hbox{$<$\kern-.75em\lower1ex\hbox{$\sim$}}}}

\def\agt{\mathrel{\raise.3ex\hbox{$>$\kern-.75em\lower1ex\hbox{$\sim$}}}}

\newcommand{\be}{\begin{equation}}
\newcommand{\ee}{\end{equation}}

\begin{document}

\title{Effect of a polymer additive on heat transport in turbulent Rayleigh-B\'enard convection} 

\author{Guenter Ahlers}
\author{Alexei Nikolaenko}
\affiliation{Department of Physics, University of California, Santa Barbara, CA 93106, USA}

\date{\today}

\begin{abstract}

Measurements of heat transport, as expressed by the Nusselt number $Nu$, are reported for turbulent Rayleigh-B\'enard convection of water containing up to 120 ppm by weight of poly-[ethylene oxide]  with a molecular weight of $4\times10^6$ g/mole. Over the Rayleigh number range $ 5\times 10^9 \alt Ra \alt 7 \times 10^{10}$ $Nu$ is smaller than it is for pure water by up to 10\%.

\end{abstract}


\maketitle

It has been known for about six decades that minute quantities of a polymer dissolved in a fluid can reduce significantly the drag experienced by turbulent flows near solid surfaces (see, for instance, \cite{SW00,PLB08}). During this time literally thousands of papers have been written about this technologically important problem. However, we are not aware of any experimental study, and know of only a single very recent theoretical investigation \cite{BCA09}, of the influence of polymer additives on turbulent convection in a fluid confined between horizontal parallel plates and heated from below (known as Rayleigh-B\'enard convection or RBC). In this letter we present data showing that the addition of about 100 parts per million (ppm) by weight of poly-[ethylene oxide] (PEO) can reduce the heat transport in turbulent RBC by 10\% or more.

The reason why the influence of dilute polymers on RBC has not been studied before is easy to see. In typical flow geometries, such as pipe flow, it was found \cite{Lu69} that significant drag reduction occurs when the polymer relaxation times are comparable to or longer than the time scales of some of the fluctuations in the turbulent flow. Turbulent RBC (see, for instance, \cite{AGL09,Ah09}) generally involves only modest Reynolds numbers. Thus, even though there is a continuum of time scales and eddy sizes, the natural time scales of the most abundant eddies or fluctuations are typically of order a second or longer,\cite{AS99,ZX01,LX10} and fluctuation times comparable to typical high molecular weight polymer relaxation times (see, for instance, \cite{VMMSM67}) are expected to be virtually absent. In addition, the viscous boundary layers (BLs) above the bottom and below the top plate, although fluctuating, are laminar because the {\it shear} Reynolds number, defined in terms of the BL thickness, usually is only of order $10^2$ or less. Thus there was no obvious reason to study the influence of minute quantities of polymers in this system. 

In view of the above discussion of the time scales of the problem it seems likely that the mechanism for the experimentally observed reduction of the heat transport in RBC is quite different from the usual drag-reduction mechanism. Since the heat transport at large Rayleigh numbers is dominated by the emission of plumes from the thermal boundary layers, the experiments suggest that the emission is reduced by the polymer additive. In view of the fact that the thermal BLs are, roughly speaking, marginally stable, and that plume emission may be regarded as a manifestation of this near-instability, it is not unreasonable that this process should be exceptionally sensitive to such factors as polymer concentration; but the precise mechanism involved in this suggested ``smoothing" effect is not immediately obvious. 

RBC occurs when the temperature difference $\Delta T=T_b-T_t$ between the bottom and top plates exceeds a critical value. At much larger $\Delta T$ the fluid flow becomes turbulent. 
Here we report results obtained with a cylindrical sample of aspect ratio $\Gamma \equiv D / L \simeq 1.00$ ($D$ is the diameter and $L$ the height) as a function of the Rayleigh number 
\be
Ra = \frac{\alpha g \Delta T L^3}{\kappa \nu}
\label{eq:Ra}
\ee
over the range $5\times10^{9} \alt Ra \alt 10^{11}$ ($\alpha$ is the isobaric thermal expansion coefficient, $g$ the acceleration of gravity, $\kappa$ the thermal diffusivity, and $\nu$ the kinematic viscosity). We used de-ionized water as the fluid, with the addition of very small amounts of PEO of nominal molecular weight $4\times10^6$ g/mole,\footnote{The PEO used was Aldrich item \# 189464.} at a mean temperature of 40$^\circ$C where the Prandtl number $Pr \equiv \nu/\kappa$ of the pure fluid is 4.38. 

We focused on the effect of the polymer additive on the convective heat transport, as expressed by the Nusselt number 
\be
Nu = \frac{Q L}{A \Delta T \lambda}
\label{eq:Nu}
\ee
 where $Q$ is the heat current passing through the sample, $A$ the sample cross-sectional area, and $\lambda$ the thermal conductivity. As the polymer weight fraction $c$ was increased from zero at constant $Ra \simeq 7\times 10^{10}$, $Nu$ decreased approximately linearly at a rate $[1-Nu(c)/Nu(0)]/c \simeq  600$.

The measurements were made  using the ``large sample" described in detail elsewhere.\cite{BNFA05} It had a diameter $D = 49.7$cm and height $L = 50.61$cm, a cylindrical plexiglas side wall of 6 mm thickness, and aluminum top and bottom plates. The plates were magnaplated\footnote{General Magnaplate California, 2707 Palma Drive, Ventura, CA 93003.} to prevent corrosion and to reduce the deposition of polymer from the solution. 

As shown elsewhere, the finite conductivity of the top and bottom plates slightly reduces the heat transport by the turbulent fluid.\cite{Ve04,BNFA05} For the present purpose the uncorrected measurements of $Nu$ are reported. This seems appropriate since the change due to the polymer addition is of interest and since the plate effect has been reported in detail before.
 
Measurements were made at constant applied temperature difference with the mean temperature $T_m$ fixed at 40$^\circ$C. The maximum applied $\Delta T$ was 10$^\circ$C, corresponding to a Rayleigh number $Ra = 6.7\times 10^{10}$. We measured the shear viscosity $\eta$ of the polymer solutions with an Ostwald viscometer and found no change due to the addition of the polymer. Thus $Ra$ was calculated using the properties of pure water at $T_m$.

\begin{figure}    
\centerline{\includegraphics[width=3.25in]{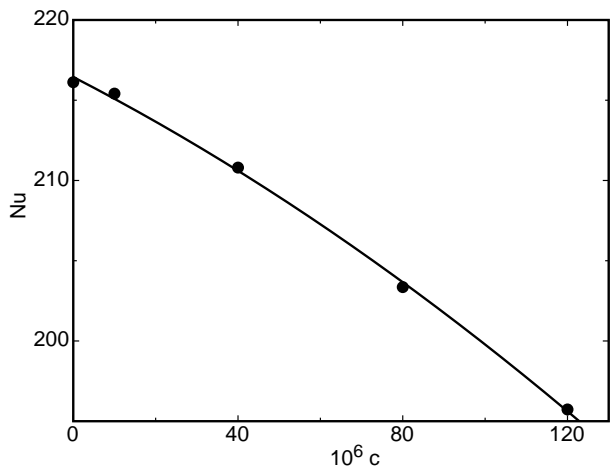}}
\caption{The Nusselt number $Nu$ as a function of the polymer weight fraction $c$ for $Ra = 6.74\times 10^{10}$. The line is a fit of a quadratic equation to the data.}
\label{fig:Nu_of_c}                                       
\end{figure}

 \begin{figure}          
\centerline{\includegraphics[width=3.25in]{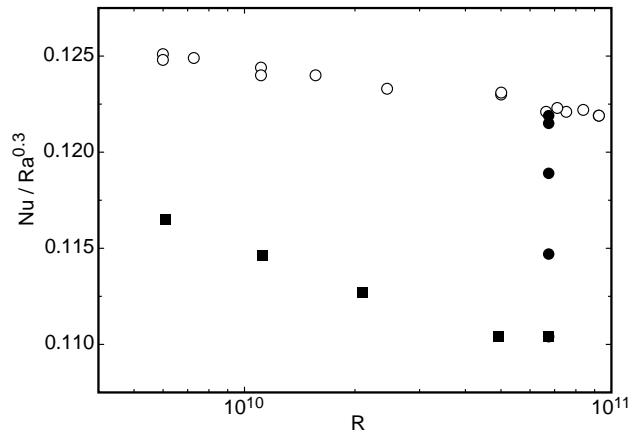}}
\caption{The reduced Nusselt number $Nu/Ra^{0.3}$ as a function of the Rayleigh number on a logarithmic scale. Open circles: pure water. Solid circles: $Ra = 6.74\times 10^{10}$ and varying polymer concentration (see Fig.~\ref{fig:Nu_of_c} for a different representation of the same data). Solid squares: 120 ppm polymer as a function of $Ra$. }
\label{fig:Nured_of_Ra}                                       
\end{figure}

In Fig.~\ref{fig:Nu_of_c} we show the Nusselt number as a function of the weight fraction of PEO. The solid line is the result of fitting a quadratic equation to the data. It is given by
\be
Nu(c)/Nu(0) = 1 - 611c ( 1 + 2600 c)\ .
\ee
The observation that the dimensionless coefficients in this equation are of order $10^3$ rather than of order unity reflects the extraordinary sensitivity of this system to the polymer additive.

In Fig.~\ref{fig:Nured_of_Ra} we show all of our measurements in the form of the reduced Nusselt number $Nu/Ra^{0.3}$ as a function of $Ra$ (using $Nu/Ra^{0.3}$ rather than $Nu$ itself has the advantage of being roughly constant and thus of allowing greater resolution). The open circles are the results for the pure fluid. The solid circles at $Ra = 6.74\times 10^{10}$ are the same data as those shown in Fig.~\ref{fig:Nu_of_c}. The solid squares are the results for $c = 1.2\times 10^{-4}$ as a function of $Ra$. One sees that the reduction of $Nu$ due to the polymer decreases slightly with decreasing $Ra$.

As mentioned in the Introduction, it is surprising at first sight that small amounts of polymer additive have any influence at all on RBC. In order to gain some insight into what {\it may} be happening, one needs to consider the major features of this system. The temperature difference is sustained mostly by two laminar, albeit fluctuating, thermal boundary layers (BLs), one below the top and the other above the bottom plate, with the interior of the system nearly isothermal in the time average but with vigorous temperature and velocity fluctuations.  When the BLs are averaged over time and in the horizontal plane, their thickness is given by $\lambda_T = L/(2 Nu)$.\cite{AGL09} In our experiment, with $L \simeq 500$ mm, $Nu$ varied approximately from 100 to 250 and thus $\lambda_T$  varied from about 2.5 to 1 mm as $Ra$ changed from $5\times 10^9$ to $10^{11}$. A useful cartoon of the system is one in which, roughly speaking, the BLs adjust their thickness so that they are marginally stable, {\it i.e.} their Rayleigh number, based on Eq.~\ref{eq:Ra} but using a temperature difference equal to $\Delta T/2$ instead of $\Delta T$ and a length $\lambda_T$ instead of $L$, approximately has its critical value for the onset of convection. Under these conditions one would expect excitations of local flow fields to occur in the BLs at irregular temporal and spatial intervals as perturbations act upon the BLs. These excitations are emitted from the BLs and have  become known as ``plumes". It seems likely that the polymer additive reduces the perturbations acting upon the BLs and thereby reduces plume emission. This would result in a reduction of the convected heat transport by the system.

Although both experiment and theory have been lacking heretofore,
very recently\cite{BCA09} a theoretical investigation, using both direct numerical simulation and analytic model calculations, was performed for a RBC system with periodic boundary conditions along the top and bottom of the sample.  In this system, which seems difficult to achieve experimentally, there are no boundary layers and all the dissipation occurs in the bulk. It was found that the heat transport is {\it increased} significantly by the addition of polymers, which is opposite to what we found experimentally in the physical system. Since there are no rigid boundaries, the theoretical result clearly is unrelated to drag reduction in the usual sense. The same authors also discuss qualitatively the more realistic case of rigid boundary conditions at the top and bottom of the sample, and suggest that either an increase or a decrease of the heat transport can be associated with polymer addition; which case occurs depends on the values of relevant parameters. However, a direct comparison with experiment does not seem possible at this time. 

In view of the discovery reported in this Letter, a great deal more work is warranted on this system. Measurements of the reduction of $Nu$ as a function of the molecular weight of the polymer will be illuminating. Optical observations of the thermal boundary layers, and of the plume density\cite{FBA08} as a function of polymer concentration, will be instructive. Determinations of the Reynolds number of the large-scale circulation in this system should be revealing. At least for now these investigations unfortunately are beyond the possible scope of this work.
 
This work was supported by the U.S National Science Foundation through Grant DMR07-02111.

\end{document}